\documentclass[reprint,superscriptaddress,
amsmath,amssymb,aps,floatfix]{revtex4-2}

\usepackage{natbib}

\usepackage{graphicx}
\usepackage{array,multirow,makecell}
\newcolumntype{C}[1]{>{\centering\arraybackslash }b{#1}}

\usepackage{url}
\usepackage{hyperref}

\newcommand{\pe}{\! = \!}
\newcommand{\thei}{\Theta^i}
\newcommand{\thef}{\Theta^f}

\begin{document}

\title{The Unexpected Dewetting during Growth of Silicene Flakes with Dendritic Pyramids}

\author{Kejian Wang} 
 \affiliation{Sorbonne Universit\'e, Centre National de la Recherche Scientifique, Institut des NanoSciences de Paris, INSP \\ 
4 place Jussieu, 75005, Paris, France}
 
\author{Mathieu Abel} 
\affiliation{%
Aix Marseille Universit\'e, Centre National de la Recherche Scientifique, Universit\'e de Toulon, IM2NP\\
Avenue Escadrille Normandie Niemen, 13397, Marseille, France
}%

\author{Filippo Fabbri} 
\affiliation{NEST, Istituto Nanoscienze – CNR, Scuola Normale Superiore \\ 
Piazza San Silvestro 12, 56127, Pisa, Italy
}%

\author{Mathieu Koudia} 
\affiliation{%
Aix Marseille Universit\'e, Centre National de la Recherche Scientifique, Universit\'e de Toulon, IM2NP\\
Avenue Escadrille Normandie Niemen, 13397, Marseille, France
}%

\author{Adrien Michon} 
\affiliation{%
CRHEA, Université C\^ote d’Azur, CNRS \\
Valbonne 06560, France
}%

\author{Adam Hassan Denawi} 
\affiliation{LPICM, CNRS, \'Ecole Polytechnique, IP Paris \\ 
91128 Palaiseau, France
}%

\author{Holger Vach} %
\affiliation{LPICM, CNRS, \'Ecole Polytechnique, IP Paris \\ 
91128 Palaiseau, France
}%

\author{Isabelle Berbezier} 
\affiliation{%
Aix Marseille Universit\'e, Centre National de la Recherche Scientifique, Universit\'e de Toulon, IM2NP\\
Avenue Escadrille Normandie Niemen, 13397, Marseille, France
}%

\author{Jean-No\"el Aqua}
\email{aqua@insp.jussieu.fr}
 \affiliation{Sorbonne Universit\'e, Centre National de la Recherche Scientifique, Institut des NanoSciences de Paris, INSP \\ 
4 place Jussieu, 75005, Paris, France}

\date{\today}

\begin{abstract}

Silicene growth on graphene has emerged as a novel method for fabricating silicon-based van der Waals heterostructures.
However, the silicene flakes produced in this manner are the result of an exotic growth mode characterized by metastable nanostructures with varying degrees of deviation from equilibrium, with large two-dimensional flakes surrounded by a rim  that coexist with small 3D islands, and, at large deposits, thick dendritic pyramids separated by a denuded zone. 
In order to rationalize and control this growth, a model is derived that revisits the dewetting thermodynamics and considers generally ignored adsorption and step-edge energies.
The model is investigated using kinetic Monte-Carlo simulations and mean-field rate equations, and implemented by close inspection of microscopy images.
This model perfectly reproduces the experimental outcomes, unveiling an anomalous growth mode, and provides guidelines on experimental conditions for high-quality silicene growth.

 \end{abstract}

\maketitle


\textbf{Introduction}

Two-dimensional (2D) materials have remarkable electronic, optical or spintronic properties \cite{NovoGeim04,NovoGeim05,ZhanTan05,TombJozs07,Geim11,Novo11,AvsaOcho20,ZhanGong21,ShanTan23,ElahKahn24,KumaSing24}. Their exfoliation/transfer has opened up a fertile field of study for different materials 
\cite{WataTani04,DeanYoun10,GeimGrig13,WangWang22,LeePark22,VergTiwa24,HaoZhan25}. 
However, their synthesis by epitaxy, which is more paradigmatic, should be better controlled and scalable up to full wafer scale. 
Its theoretical description is challenging due to its complexity, multi-scale nature and dependence on atomic processes \cite{MomeJi20}. It requires significant development in order  to guide experiments in the search for optimal growth parameters. 
This is the case in particular for silicene that is predicted to be a Dirac materials with a non-negligible band-gap promising for overcoming graphene's intrinsic limitations \cite{TakeShir94,CahaTops09,OughEnri15}, but that cannot be produced by exfoliation. Numerous growth experiments were conducted, yielding variable degrees of success in terms of reality, quality, extent of the flakes, and potential real-life applications   \cite{LalmOugh10,VogtDePa12,FengDing12,ChenChen12,BernLeon12,MengWang13,BernLeon13,LinAraf13,BernBore15,DeCrBerb16,SattLaco18,ColoFlam21,BenJAbel22,MassPrev23}. Most of the deposition on metals showed strong interactions with their substrate. A natural alternative is to grow silicene on van-der-Waals templates for decoupling the 2D adlayers from their substrates. It is also promising as the fabrication of functional devices requires non-metallic wafers compatible with the semiconductor industry. The recent epitaxial growth of Si on Highly Oriented Pyrolytic Graphite \cite{DeCrBerb16} and on Graphene(Gr)/SiC \cite{BenJAbel22} have demonstrated the epitaxy of free-standing 2D flakes but with rather small sizes. An initial simplified modelization was used to rationalize the growth of small faceted crystallites based on the dynamics of dewetting in near-equilibrium conditions. 

The anomalous growth of silicene on Gr grown by CVD on a 6H-SiC(0001) substrate \cite{JabaMich14,BenJBerb21} is atypical. Firstly, it results in unique morphologies consisting of large compact and irregular flakes with lateral dimensions of up to 200\,nm, which are neither facetted nor dendritic, and surrounded by a thicker rim. Secondly, these flakes coexist  with 3D dendritic islands, 3 to 4 monolayers thick, which grow simultaneously. The coexistence of morphologies with such differences corresponds to different levels of deviation from equilibrium.
To rationalize this unique and anomalous growth mode, we derive a model incorporating atomic processes at step edges and dewetting thermodynamics. We solve the far-from-equilibrium many-body dynamics by kinetic Monte Carlo simulations and mean-field equations. 
We develop an original approach that allows microscopic parameters to be deduced from a detailed analysis of the morphological characteristics revealed by microscopy experimental images. Based on this analysis, we are able to replicate the experimental growth mode, and its long-term evolution leading to dendritic pyramids separated by a denuded zone.
The model's universality of ingredients and capacity for reproducing complex morphologies provide a novel framework for explaining 2D crystal growth well beyond the van der Waals epitaxy of silicene on Gr.

\textbf{Experiences} 

\begin{figure*}[ht] \centering
\includegraphics[width=7.cm]{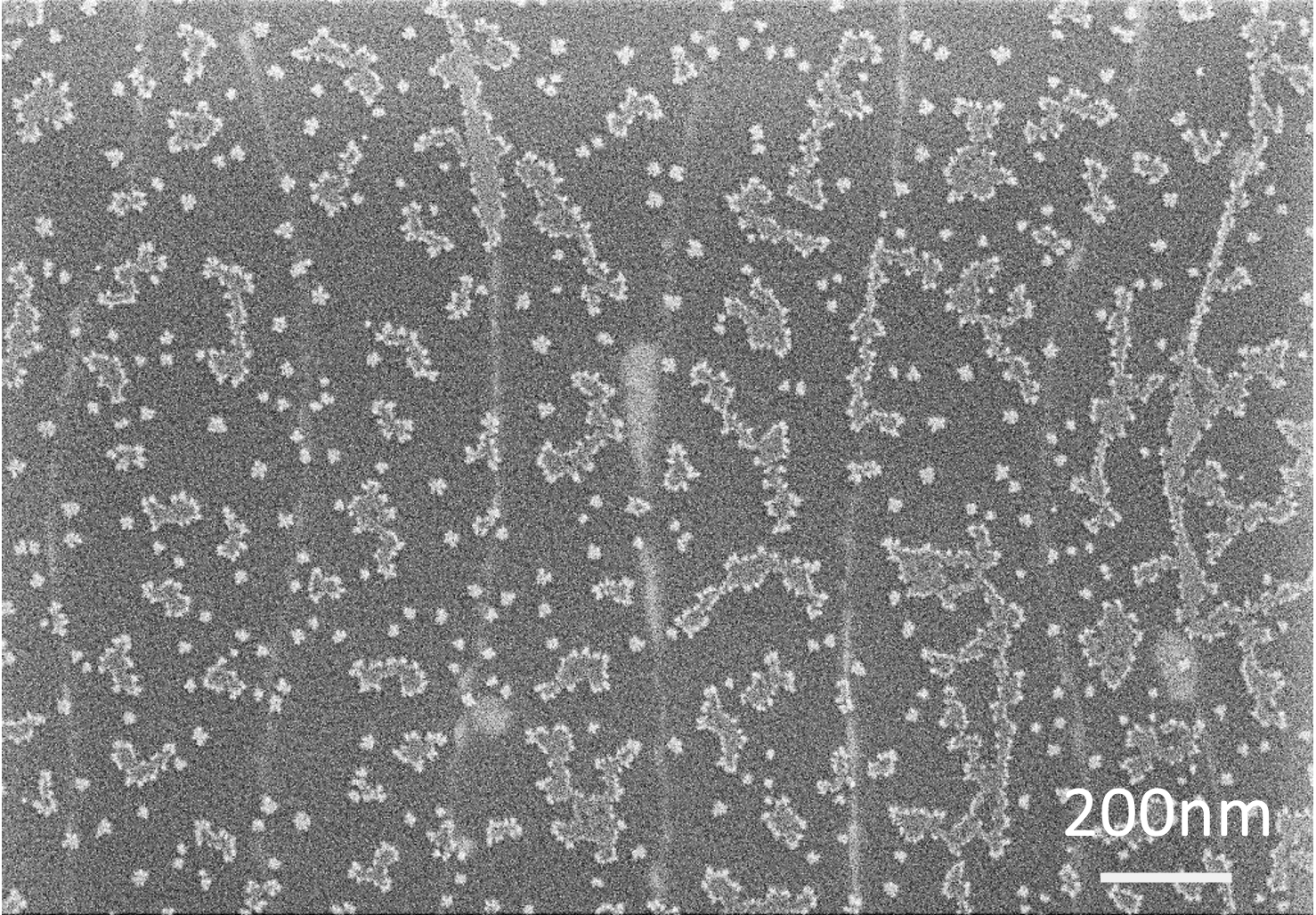}
\includegraphics[width=5.7cm]{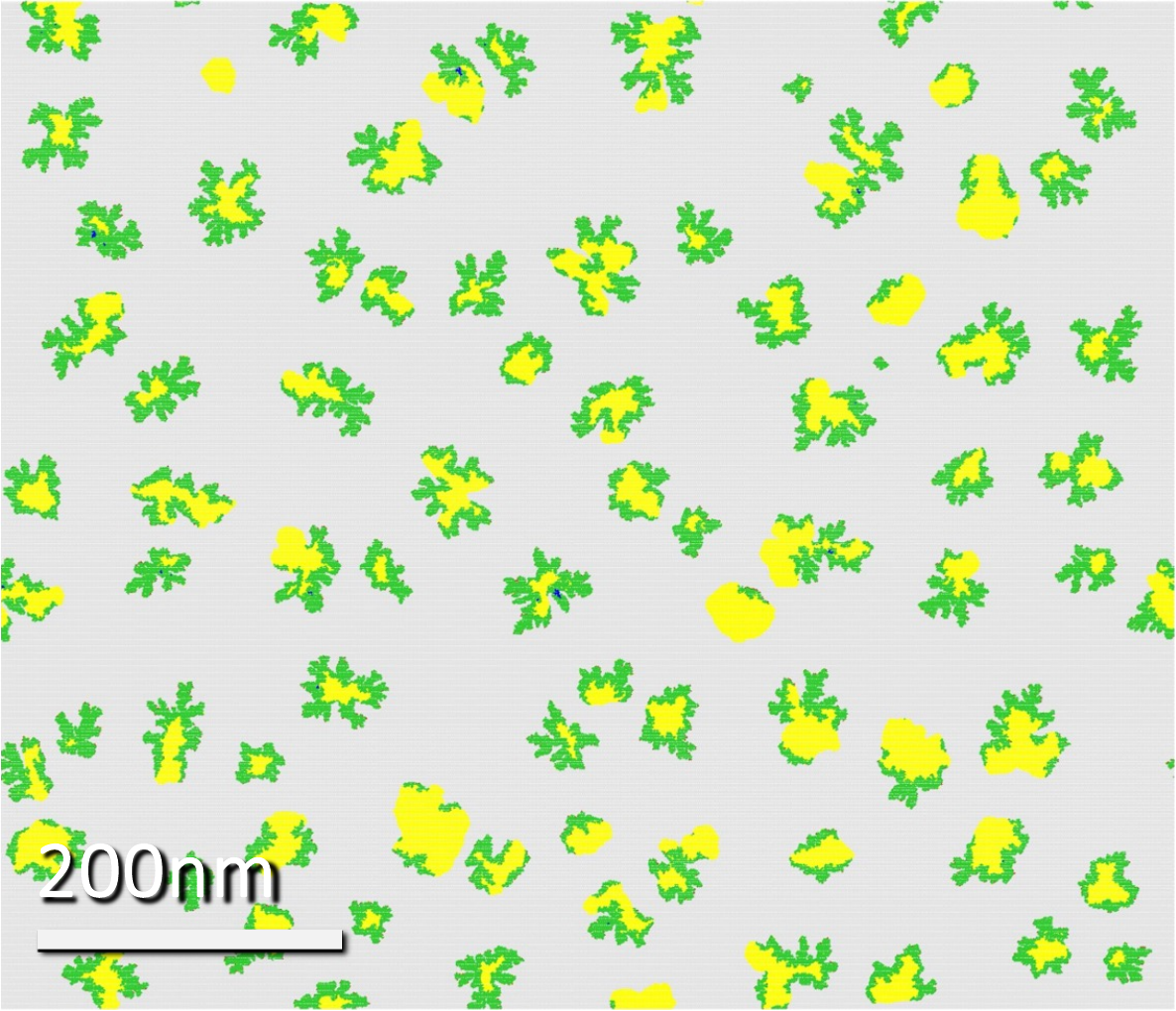}
\includegraphics[width=1.cm]{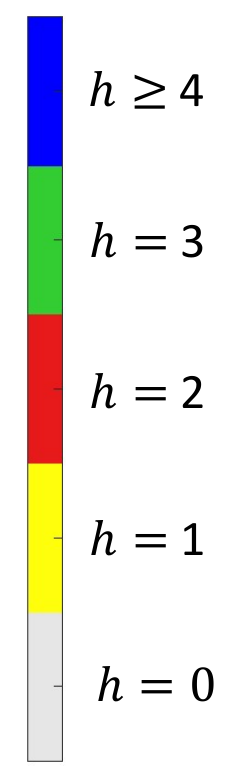}
\caption{a) SEM images of the silicene surface after deposition at $T \pe 300$\,K and $F\pe 0.18$\,ML/h of $\thei \pe 0.35$\,ML. (b) Kinetic Monte-Carlo simulation of the model for the epitaxy of Si on Gr for the experimental parameters and the simulation parameters given in Table \ref{param}. The film of the growth is available online \cite{filmKMCSiGr}. The box size is $2000\times 2000$ in lattice constant.
}
\label{figexp1}
\end{figure*}

\begin{figure*}[ht] \centering
\includegraphics[width=7.cm]{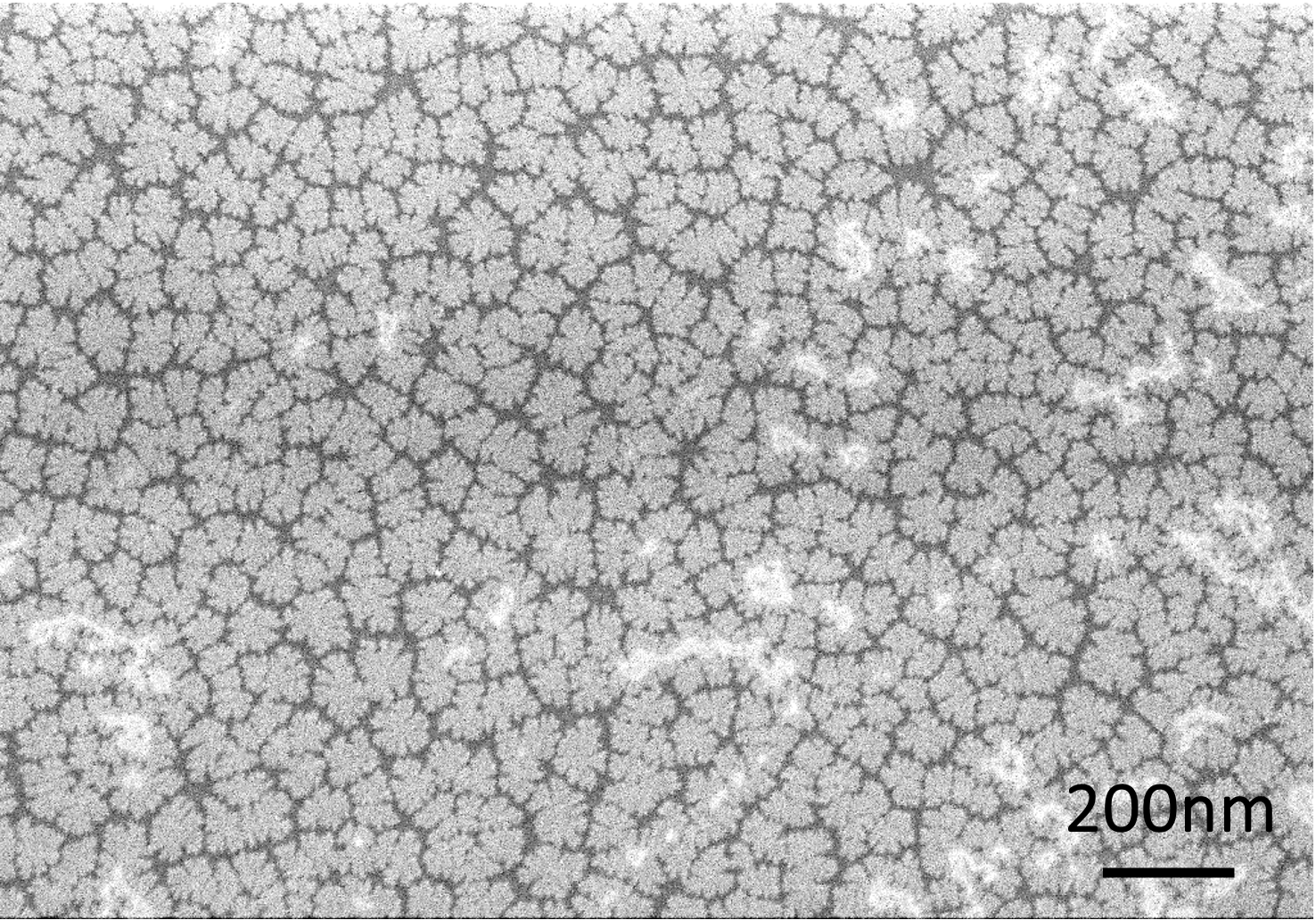}
\includegraphics[height=4.87cm]{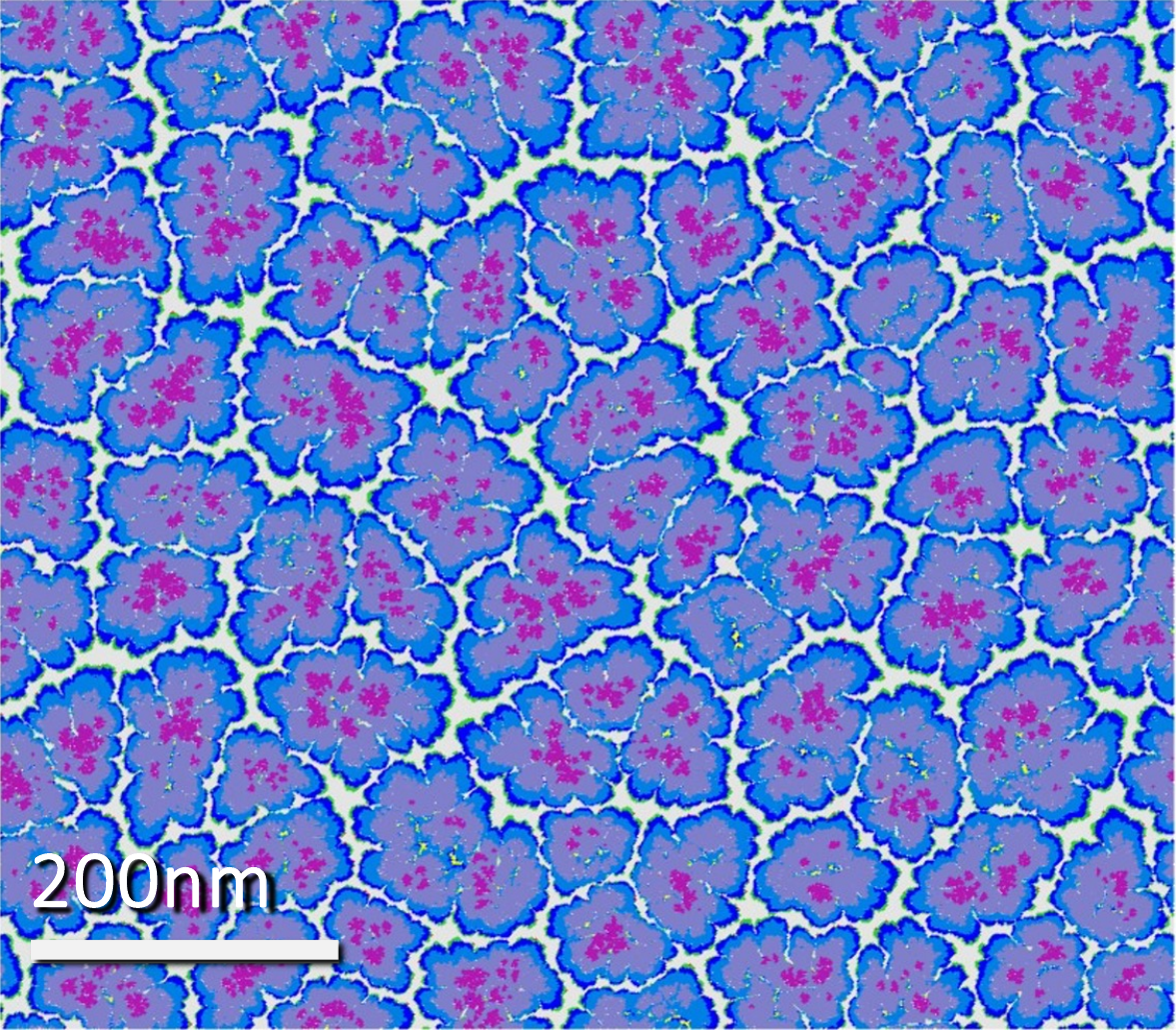}
\includegraphics[height=4.85cm]{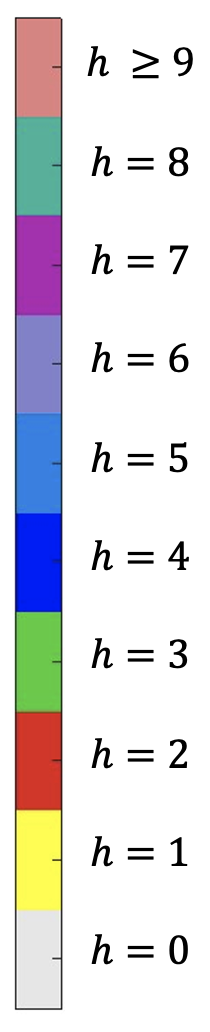}
\caption{Same Figure as \ref{figexp1} with $\thef \pe 4.2$\,ML. }
\label{figexp2}
\end{figure*}

Silicon was deposited on Gr on SiC. Graphene was grown by CVD under hydrogen \cite{BenJBerb21} on the Si-face of N-doped 6H-SiC in conditions leading to the formation of a uniform monolayer (ML) lying on a buffer layer \cite{MichLarg12}. Prior to Si deposition, Gr/SiC wafers were cleaned in situ by heating to 900$^\circ$\,C under ultra-high vacuum to remove air contamination. The cleanliness of the 1\,ML Gr buffer layer was assessed by LEED, XPS and STM (VT-Omicron). Silicon was deposited at room temperature on the ML Gr template layer from a direct-current evaporation source. The deposition rate and surface cleanliness deposition were monitored by XPS with an Omicron EA 125 energy analyzer and a monochromatized Al X-ray source. Coverages of $\thei \pe 0.35$\,ML and $\thef \pe 4.2$\,ML were obtained at a constant flux $F\pe 0.18$\,ML/h. The silicene deposits were characterized by SEM and Micro-Raman spectroscopy using a Renishaw Invia Qotor equipped with a confocal optical microscope. A 532\,nm excitation laser was employed with an excitation power of 0.5\,mW and a 1800 line per millimeter grating (spectral resolution 2\,cm$^{-1}$). The laser beam was focused onto the sample by a x100 objective with numerical aperture $NA\pe0.85$ and a spot size of 800\,nm.

For $\thei$, the SEM images in Fig.~\ref{figexp1} first exhibit large monolayer silicene flakes. 
These flakes consist of two parts: a flat central area with a thickness of $\sim 0.35 \pm$ 0.02 nm and lateral dimensions about 100–200 nanometers, and a periphery characterized by a 3\,ML rim which may encircle either part or all of the flake. Raman peaks representative of silicene have also  been evidenced. Two new peaks appeared: one at 560\,cm$^{–1}$, which is interpreted as the zone-center $E_{2g}$ vibrational mode of Si-ene (equivalent to the so-called G-like peak of Gr), which was predicted by theoretical studies at 570\,cm$^{–1}$ (34) and another at 240\,cm$^{–1}$, which is assigned to the breathing mode. We detected the Raman peaks after several months of exposure to air, demonstrating their robustness which is an important feature \cite{LiuLei14,BenJAbel22}. 
The silicene flakes are compact and irregular in shape, neither dendritic (as found in highly out-of-equilibrium conditions) nor faceted (as found close to equilibrium). They coexist with numerous and small 3D islands. Eventually, all these objects transform into dendritic pyramids as growth proceeds, see Fig.~\ref{figexp2}. For $\thef$, one observes a dense collection of 3D dendritic pyramids still separated by a denuded zone. This transformation suggests that 3D objects are more stable, and that 2D flakes would be intermediate states.
Finally, a quantitative analysis of microscopy images reveals parameters useful for modeling purposes. First, the island density is $\rho \!\simeq \! 150\,\mu$m$^{-2}$. Second, we analyse the layer coverage $\theta_{n}$ (i.e.~the total number of atoms on the $n$-th level divided by the number of sites) with $\theta_{h \geq 2} \pe \sum_{n = 2}^\infty \theta_{n}$. For $\thei$, we find $\theta_{h = 1} - \theta_{h \geq 2} \pe 0.072$ and $\theta_{h \geq 2} \pe 0.094$ \footnote{they satisfy $\theta_1 + 3*\theta_{h \geq 2} \pe \thei$, assuming a 3 monolayer thick rim}.


\textbf{Modelization} 

The growth mode thus described is unique and does not correspond to the standards of crystalline growth \cite{PimpVill98,Mark03,EinaDiet13}. Of special concern is the coexistence of distinct morphologies, 2D flakes and 3D fractals, which are characterized by varying degrees of deviation from equilibrium. A dewetting model in \cite{BenJAbel22} described a single island with a rim in a state of near equilibrium with a faceted shape. However, it does not account for irregular two-dimensional islands that are further from equilibrium, or three-dimensional fractals, or the coexistence of these shapes. We propose here a model coupling van der Waals epitaxy and dewetting, based on a reexamination of surface thermodynamics. We take up the classical modelization of reversible aggregation in the presence of deposition, diffusion and attachment/detachment \cite{RatsZang94,PimpVill98}. We consider a lattice model with honeycomb lattices arranged in sheets stacked on top of the other \cite{*[{neglecting higher-order geometric effects, see e.g.~ }] [{}] BhuiTera21}. We restrict jumps to a maximum of one atomic height. The energy barriers for atomic processes are generally broken down into $E_S + n_i E_N$. For an atom $i$ at the top of an atomic column, the diffusion energy barrier $E_S$ (respectively $E_N$) results from interaction with the underlying layers (respectively with the $n_i$ in-plane nearest neighbors). This solid-on-solid description is used for the epitaxy of different materials (Ge/Si, Si/SiO$_2$, Si/Ag, Cu, GaN, CdTe, graphene, silicene, MoS$_2$, WS$_2$, WSe$_2$ etc), with different effects (elasticity, wetting, reconstruction etc)
\cite{MeixScho01,LamLee02,RussSmer06,ZhuPan07,AquaFris08,Lam10,RabbWorm09,BussChey11,SchuSmer12,LinHamm12,GailAqua13,GhosRang14,GailChan15,JianHou15,EnstBrom16,GoviWarn16,NieLian16,YueNie17,ChugRang17,LiZhan19,WuYang19,ToAlme21,CheiTo21,KongZhua21,WangPrev24}. 
In heteroepitaxy, the atomic environment depends on the number of film layers, so that energy barriers should depend on the atomic height $h$, describing wetting effects. Yet, in order to automatically ensure detailed balance \cite{RussSmer06,BenJAbel22}, a common implicit assumption is that the barriers $E_S$ and $E_N$ are the bond energies, implicitly implying that atoms detach from the crystal during each elementary process. 

\begin{figure}[ht] \centering
\includegraphics[width=8.5cm]{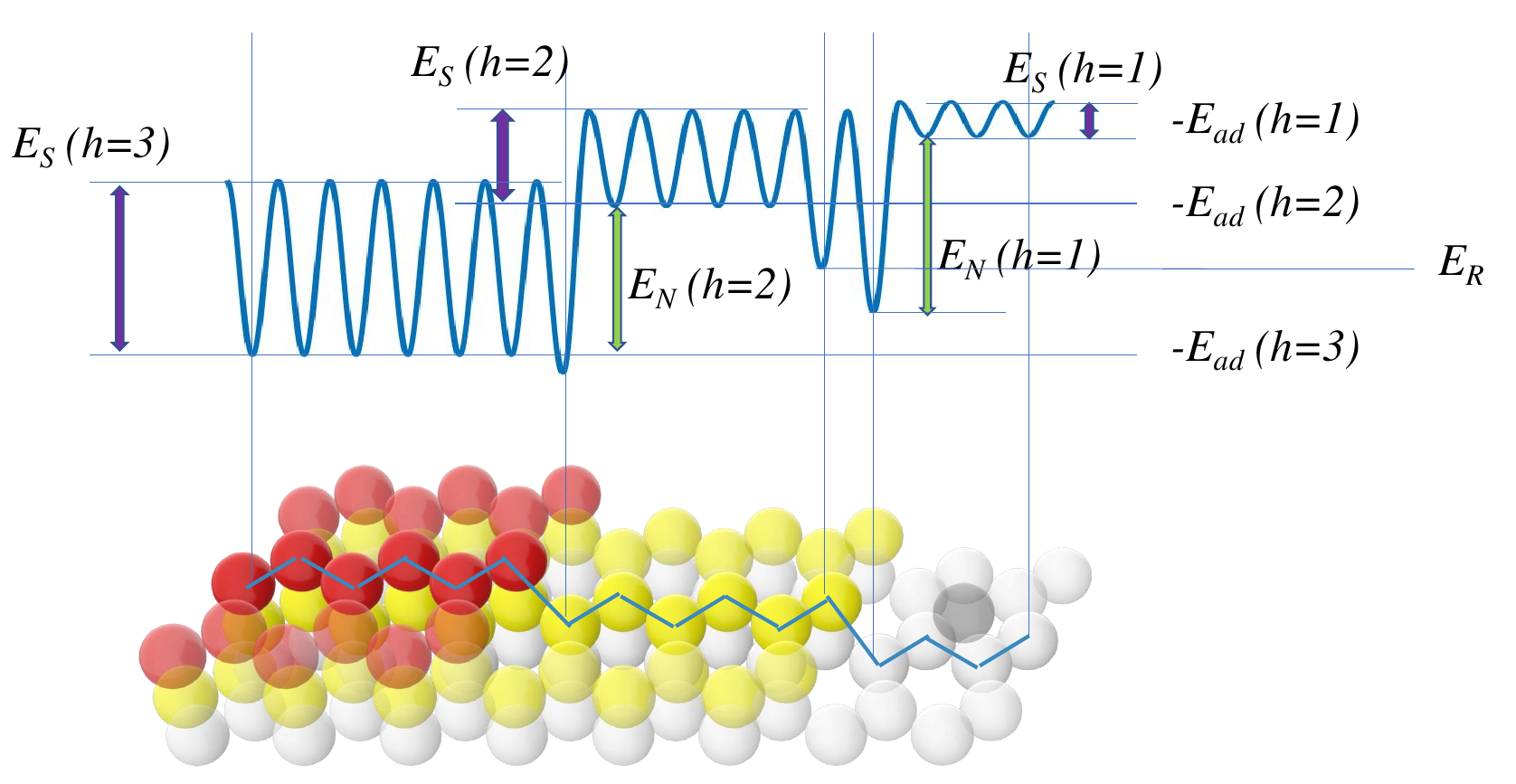}
\caption{Energy path experienced by a test atom (in dark gray) diffusing on the substrate (light gray) and a flake (with yellow ($h \pe 1$) and red ($h\pe 2$) atoms).}
\label{figNRJ}
\end{figure}

We go back on this overly strong assumption by introducing the adsorption energy $E_{ad}$ defined as the difference between the total energy of an adatom-surface system and that of the adatom and surface in isolation. Then, the energy of an atom $i$ is $-E_{ad}(h_i) - n_i E_N(h_i)$, while the saddle point of an atom diffusing at constant $h$ is $-E_{ad}(h) + E_S(h)$, see Fig.~\ref{figNRJ}. Similarly, the saddle point of an atom transitioning up or down a step edge between heights $h$ and $h+1$ is $-E_{ES}^{h/h+1}$, potentially describing Ehrlich-Schwoebel effects \cite{JeonWill99,PimpVill98}. The energy barrier for an elementary process where an adatom initially at $h_i$ ends up at $h_f$ is then
\begin{multline} 
\label{DE}
\Delta E = E_S(h_i) + n_i E_N (h_i)  \hspace{2.cm} \mbox{if $h_i \pe h_f$,}\\
  = -E_{ES}^{h_i/h_f} + E_{ad}(h_i) + n_i E_N (h_i)   \, \, \,   \mbox{if $h_i \! \neq \! h_f$,}
\end{multline} 
still consistent with detailed balance. In this case, differences in adsorption energy regulate the kinetics, controlling the flow of adatoms at step edges, and we argue that they concern all systems where wetting is at work.
In addition, the presence of flakes rims suggests the presence of a stabilization at the flakes periphery. It may be incorporated by a decrease by a value $E_R$ in the energy of the position at the top edge of a step, see Fig.~\ref{figNRJ}. Atoms stabilized in this way are simply characterized by $h \pe 2$ with at least one neighbor at $h \pe 0$ and one at $h \pe 1$. This effect is pivotal to accurately reproduce experiments; nevertheless it is the balance between this effect and those associated with dewetting (in terms of diffusion barriers and adhesion energy), that facilitates the reproduction of  experimental morphologies, see Supplementary Materials. 
This effect was inferred here based on a detailed comparison between theory and experiment. A priori, a boundary can be associated with excess quantities. Since the step edge is the boundary of the terrace, thermodynamic quantities such as adsorption energy can vary in its vicinity, due to local inhomogeneities e.g. in dangling bonds and surface reconstruction. $E_R$ may then be viewed as the excess of the adsorption energy. Similarly, edge energies are sometimes included in coarse-grained thermodynamic descriptions. In terms of energy barriers, the variations described by Erlich-Schwoebel barriers are also due to local variations in thermodynamic quantities in the vicinity of a step. 

We analyze the resulting growth dynamics through kinetic Monte Carlo simulations. We use the rejection-free $N$-fold algorithm that avoids discarded attempts \cite{BortKalo75,DeViSand05}. The event catalog includes deposition with a flux $F$ and atomistic events with a rate $\nu_0 \exp(-\Delta E / k_B T)$, where $\nu_0 \pe \alpha k_B T / h$ is the attempt frequency with the prefactor $\alpha$. In order to precisely describe the experiments, we needed a description of wetting with at least three layers, $h \pe 1$ (first deposited layer), $h \pe 2$ and $h\! \geq \!3$. In the following, we simplify the Ehrlich–Schwoebel effect by selecting a saddle point for the transition between two layers aligned with the one on the upper layer, $-E_{ES}^{h/h+1} \pe -E_{ad}(h+1) + E_S(h+1)$. The model is thence parameterized by nine parameters, $E_S$, $E_N$ for $h\pe 1, 2$ and $\geq \!3$, $E_R$, $E_{ad}(2)-E_{ad}(1)$ and $E_{ad}(3)-E_{ad}(2)$. 

These values are adjusted considering the information provided by the microscopy images, analyzing the islands shapes, presence of a rim, its continuity and height, the coexistence of various morphologies etc. $E_S(h\pe1)$ and $\alpha$ that regulate island nucleation are first set thanks to the island density and ab initio calculations \footnote{For $E_S(h\pe1)\pe 0.01$\,eV, $\alpha \pe 1.25\,10^{-4}$ gives a density of $200\,\mu$m$^{-2}$, while $\alpha \pe 5\,10^{-4}$ gives a density of $110\,\mu$m$^{-2}$}. 
Using density functional theory within the VASP code \cite{PerdZung81,KresFurt96} and incorporating the van der Waals dispersion correction via the  GGA+vdW method \cite{Grim06}, we investigated the diffusion of a Si atom on a monolayer graphene. This barrier is found to be very small, 0.011\,eV, enabling Si to move easily between two bridge sites \cite{longSiGr}.
$E_N(h\pe 1)$ is in turn determined with the flakes morphologies that are experimentally rather irregular, neither hexagonal (equilibrium shape at low $E_N$), nor dendritic (irreversible shape at large $E_N$ \cite{RatsZang94}). Moreover, ab initio calculations show that the adsorption energy of Si on Gr is $E_{ad} \pe 1.30$\,eV \cite{AktuAtac10}. For higher layers, Si on top of Si layers is expected to be more strongly bonded as Si dewets Gr \cite{LupiKitz13}, so that $E_S$ is expected to increase with $h$ \footnote{this increase must be progressive to avoid dendritic islands forming on top of the flakes}. With this methodology, some parameters such as $E_{ad}$ or $E_R$ are sensitive with an uncertainty of $\pm 0.02$\,eV, while others are less critical

\begin{table}[h]
\begin{tabular}{|C{1cm}|C{2cm}|C{2cm}|C{1.5cm}|}
\hline  \rule{0mm}{3mm}  & $h\pe 1$ &  $h\pe 2$ & $h \! \geq \! 3$  \\
\hline  \rule{0mm}{3mm} $E_S$ 	& $0.01 $    &     $0.16 \pm 0.04$   &   $0.33  \pm 0.15$  \\
\hline  \rule{0mm}{3mm} $E_N$ 	& $0.50 \pm 0.02$   &     $0.38 \pm 0.02$   &     $0.4 \pm 0.15$ \\
\hline  \rule{0mm}{3mm} $E_{ad}$ & 1.30   &   $1.46 \pm 0.02$    &   $1.8 \pm 0.15 $    \\
\hline  \rule{0mm}{3mm} $E_{ES}$ & \multicolumn{2}{c|}{$1.30 \pm 0.04$} &{$1.47$}  \\
\hline  \rule{0mm}{3mm} $E_R$ & \multicolumn{3}{c|}{$0.14 \pm 0.02$}  \\
\hline 
\end{tabular}
\caption{Parameter set of the KMC simulations, with the prefactor $\alpha \pe 2.5\,10^{-4}$. }
\label{param}
\end{table}


\textbf{Results and discussion}

Following incremental variation, the set of parameters given in Table~\ref{param} allows a good comparison between simulations and experiments. The movie of the film's growth for $T\pe 300$\,K and $F\pe 0.18$\,ML/h is available online \cite{filmKMCSiGr}. The morphologies obtained for $\thei$ and $\thef$ are shown in Fig.~\ref{figexp1} and \ref{figexp2}. 
We configured the model using a back-and-forth approach in order to reproduce most of the singular facts of the anomalous growth mode. The constraints we imposed are both qualitative, in terms of comparing the typical morphologies of simulations and experiments as explained below, and quantitative (concerning island density, rim height, or proportion of atoms at 1 or 3ML). Indeed, for $\thei$, the simulation can reproduce large 2D flakes with either a continuous or discontinuous rim at their perimeter that do coexist with smaller 3D dendritic islands. Both the flakes rim and 3D dendrites are 3\,ML thick. The inner edge of the rim is fairly smooth, while the outer edge may have dendritic growths, as observed in microscopy images. For $\thef$, the simulations also reproduce that all the islands have been transformed into dendritic pyramids, similar to wedding cakes with dendritic layers, 5 to 6\,ML thick. These pyramids do not fully coalesce and a thin, denuded zone separates them from each other. Semi-quantitatively, the simulations reproduce flakes up to 60\,nm in width considering the lattice parameter of free silicene $a\pe 0.385$\,nm \cite{TakeShir94}. The island density for $\thei$ is $140$\,$\mu$m$^{-2}$ close to the experimental value ($150$\,$\mu$m$^{-2}$). Finally, the surface coverages of both the monolayer areas and 3D ones are $\theta_{h \pe 1} - \theta_{h \geq 2} \pe 0.082$ and $\theta_{h \geq 2} \pe 0.091$ for $\thei$, close to their experimental counterparts ($0.072$ and $0.094$). 
A discrepancy concerns the size difference between the flakes and dendrites: even if the 3D dendritic islands are smaller than the 2D flakes in the simulation, the distribution of sizes between 3D and 2D islands in simulations is tighter than in experiments. This may be due to statistical fluctuations, step edge effects or the presence of spatial inhomogeneities in the graphene layer, which can have a significant influence on the nucleation barriers. Fluctuations and inhomogeneities are indeed expected to be important in a growth process that relies and depends on metastable states. 

The 2D silicene flake appears here as an intermediate and metastable stage in the competition between growth and dewetting. The simulations reveal that 2D islands initially nucleate and grow with a compact shape, see the movie \cite{filmKMCSiGrzoom}. 
Atoms can detach from an island and move preferentially to its upper layer due to dewetting. As the island grows, the probability of nucleation on the upper layer increases and the rim begins to develop, first spreading around the flake. Thanks to dewetting again, the upper layers attract most of the subsequent growth, giving rise to thick dendritic islands. 3D islands being more stable, the 2D flakes gradually disappear. The coexistence of the former with ringed 2D flakes results from fluctuation in the nucleation process.  

It is remarkable that all the effects mentioned above play a crucial role to finely reproduce the anomalous growth mode. Without $E_R$, certain parameter sets can lead to mass accumulation at the edges, but they produce dendritic fingers on top of the flake. 
Similarly, adsorption energies play a key role in the timing of the rim formation. A decrease in $-E_{ad}(h \pe2)$ by 0.02\,eV accelerates the atomic flow to the higher layers and all the flakes have transformed into 3D dendrites at $\thei$. Conversely, an increase by 0.02\,eV implies that the rim has barely begun to grow at $\thei$. 
Even if the parameter values are inevitably influenced by uncertainties and assumptions, we claim that our dynamical dewetting model offers a robust framework for rationalizing the experimental anomalous growth mode.
In standard modeling (see e.g.~\cite{YueNie17,NieLian17}) and in classical experiments, the growth shapes are either close or far from equilibrium. The latter ones are rather (i) flat dendrites for a sub-monolayer deposit, (ii) which grow in a layer-by-layer mode for thicker deposits. We can indeed find this classical mode with other sets of parameters. However, the shapes observed and theoretically reproduced here are (i) compact 2D flakes with a rim in the sub-monolayer regime, and (ii) thick dendritic pyramids for thick deposits. Yet, we can obtain many other far-from-equilibrium morphologies and growth modes by changing the model's parameters. Only a subtle balance between the parameters as given in Table \ref{param} allows to reproduce the experiments, with relatively minimalist and necessary ingredients (wetting effects, edge effects, adsorption energies). It highlights the complexity that can emerge from the combination of kinetic processes that are nonetheless elementary. 

\begin{figure}[ht] \centering
\includegraphics[width=8.6cm,height=5.cm]{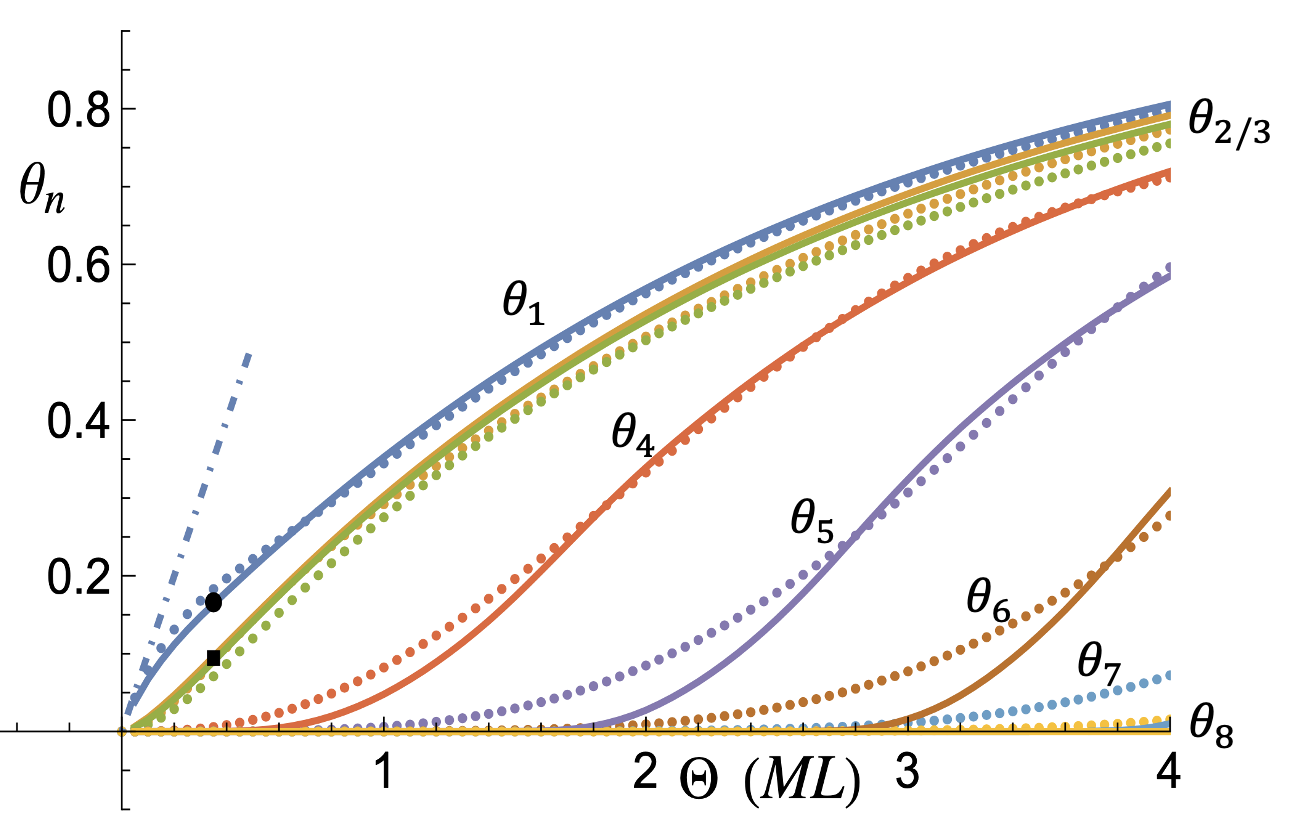}
\caption{Evolution of the layers' coverages ($n\pe 1, \ldots 8$) with the deposited height for KMC simulations (solid line), kinetic model (dotted line) and experiments (black dot and square for $\theta_1$ and $\theta_3$). The initial dot-dashed line corresponds to $\theta \pe F\, t$. The parameters are $A_1 \pe A_2 \pe 0$, $A_3 \pe 0.7$, $A_4 \pe .95$ while $A_n \pe 1$ for $n\geq 5$, $\delta \pe 0.5$, $\delta^* \pe 0.1$, $\Theta_c \pe 0.28$, $\epsilon \pe 0.03$ and $J \pe 6$.
}
\label{figTheta2h}
\end{figure}

Thanks to these results, we derive kinetic equations describing the evolution of the layers' coverages in terms of birth-death models \cite{CohePetr89,KimLee19}. In this mean-field framework, $\theta_n$ evolves thanks to the deposition flux and the transfer to the $n$-th level of a fraction $\alpha_n$ of the atoms freshly landed at the level $n+1$:   
\begin{equation} 
\label{dynamic}
\frac{1}{F}\frac{d\theta_n}{dt} = (1-\alpha_{n-1})\left(\theta_{n-1}-\theta_n\right) + \alpha_n  \left(\theta_{n}-\theta_{n+1}\right) + j_n(t) \, .
\end{equation} 
Assuming \cite{CohePetr89} that the atoms at level $n$ are evenly distributed along the steps, we get
\begin{equation} 
\alpha_n = A_n \frac{d(\theta_n)}{d(\theta_n)+d(\theta_{n+1})} \, ,
\end{equation}
with the accessible perimeter length $d(\theta) \pe \sqrt{\theta}$ if $0\leq \theta \leq 1/2$, and $d(\theta) \pe \sqrt{1-\theta}$ if $1/2\leq \theta \leq 1$. 
Moreover, $A_n$ should depend on $n$ here because of wetting effects. Yet, without $j_n$, Eqs.~\eqref{dynamic} cannot reproduce the simulations coverages as $\theta_2$ and $\theta_3$ are populating faster than the model can reproduce, and are very close together.
To solve this, we assume that dewetting induces a net flow $j_n$ from $n \pe 1$ to higher layers, with the simple choice $j_1 \pe -j_2/2 \pe - j_3/2$ while $j_{n \geq 4} \pe 0$. We consider $j_1(t)$ to be proportional to the islands perimeters $\sqrt{\theta_1}$ \footnote{assuming a constant island density which is quickly satisfied in the simulations} as the numerous attachments/detachments to the steps in the simulations suggest a situation close to local equilibrium. We finally regularize $j_1$ to ensure $\theta_n \geq \theta_{n+1}, \forall n$ and extinguish this flow when the rim spreads all around the islands after $\Theta \simeq 0.3$, 
\begin{equation} 
j_1(t) \pe J \sqrt{\theta_1(t) \, r[\theta_1(t)]} r^*[\theta_1(t)-\theta_2(t)] r^*[\theta_2(t)-\theta_3(t)] \, , 
\end{equation} 
with $r[\theta] \pe \left\{1-\tanh\left[(\theta-\theta_c)/\delta\right] \right\}/2$ and $r^*[\theta] \pe \left\{1+\tanh\left[(\theta-\epsilon)/\delta^*\right] \right\}/2$. 
The resulting evolution with appropriate parameters is plotted in Fig.~\ref{figTheta2h} and reproduces very satisfactorily the simulation and experimental data. It highlights the main processes mentioned above that govern the evolution beyond nucleation: attachment/detachment to the steps and dewetting, that quickly transfer adatoms onto the islands and that favor nucleation on the upper layers. It is noticeable that $\theta_1(t)$ deviates rapidly from its initial tangent $F \, t$ showing that the mass transfer onto the islands occurs rapidly. 

\textbf{Conclusion}

Epitaxy of Si on Gr initially yields silicene flakes with a thick rim at their periphery that are free-standing, compact, irregular and coexist with 3D islands. They eventually lead to dendritic pyramids separated by a denuded zone for thick deposits. In order to rationalize this new anomalous growth mode, we derive an original growth model that perfectly accounts for the non-standard experimental facts. It describes the interplay of van-der-Waals epitaxy with dewetting, wherein adsorption energies, which are often disregarded in the modeling of epitaxy, play a pivotal role. 
The parametrization is achieved through an original methodology that uses the precise comparison of the morphologies as revealed by microscopic imaging, with the outcomes of kinetic Monte-Carlo simulations. The silicene flakes appear as metastable states evolving towards more stable dendritic pyramids. 
The anomalous growth mode results from a complex balance between the mechanisms included in the model (dewetting, adsorption energy, step edge energy) that are well known in surface thermodynamics. The novelty lies in this combination  that makes it possible to reproduce the complex set of anomalous morphologies.
The present modeling and understanding of the growth mechanisms are the first building blocks for future theoretical predictions aimed at producing full-wafer flakes, considering e.g.~annealing, vicinal substrates or inhomogeneities in adsorption energies. We also argue that the methodology should be applicable to other epitaxial systems, in order to reveal growth mechanisms and find experimental estimates of energy barriers.


\textbf{Aknowledgements}

Funding from the French Agence Nationale de la Recherche, under the grant ANR-ComeOn-ANR-22-CE09-0012-01 is gratefully acknowledged.

\end{document}